\documentclass[print]{revtex4}
\usepackage{amsmath,amssymb}
\usepackage{graphicx}

\begin{document}

\title{Quantum state transfer from light to molecules via coherent two-color photo-association
in an atomic Bose-Einstein condensate}

\author{Hui Jing$^{1}$ and Mingsheng Zhan$^{2,3}$}

\affiliation{1. Department of Physics, The University of Arizona,
1118 East 4th Street, Tucson, AZ 85721\\
2. State Key Laboratory of Magnetic Resonance and
Atomic and Molecular Physics,\\
 Wuhan Institute of Physics and Mathematics, Chinese Academe of Science, Wuhan 430071, People's Republic of China\\
  3. Center for Cold Atoms, Chinese Academe of Science, People's Republic of China}

\begin{abstract}
By using a quantized input light, we theoretically revisit the
coherent two-color photo-association process in an atomic
Bose-Einstein condensate. Under the single-mode approximations, we
show two interesting regimes of the light transmission and the
molecular generation. The quantum state transfer from light
to molecules is exhibited, without or with the depletion of trapped atoms. \\


\end{abstract}

\baselineskip=16pt

\maketitle

\indent Since the remarkable realizations of Bose-Einstein
condensates (BEC) in cold dilute atomic gases, many novel
properties of the macroscopic quantum degenerate gases have been
experimentally exhibited or theoretically predicted [1]. Recently,
by creating a quantum degenerate molecular gases via a magnetic
Feshbach resonance [2-3] or an optical photo-association (PA)
[4-5] in an atomic BEC, the appealing physical properties of the
formed atom-molecule mixtures have attracted much interests as a
whole new dimension in the study of ultracold atomic physics and
even atom optics [2-5]. The Bose-enhanced coherent PA process was
also termed as superchemistry since it is completely beyond the
classical Arrhenius chemical kinetics in such an ultralow
temperature of the $\mu k$ range [6]. Although many novel quantum
effects are predicted for the coupled atomic-molecular
condensates, such as the molecular damping due to the quantum
noise terms [6], the essential features of the current PA
experiments can be well described by the single-mode or even
mean-field approaches, such as the famous experiment of coherent
two-color PA experiment of Winkler et al. [4]. Hence the extremely
complicated methods fully taking into account of the dissipated
effects are $not$ practically needed.

On the other hand, there also have been many interests recently in
making and exploring the new applications of a coherent atomic
beam or an atom laser [7]. Most recently, Haine et al. introduced
a scheme to generate controllable atom-light entanglement by using
a Raman atom laser system [8]. The key point of their scheme is to
use an input squeezed light instead of a classical light [9].
Therefore an interesting question may arise: is it possible to
realize the quantum control of molecule-light system or, as the
first step, a non-classical molecule laser? The possible novel
applications of a molecule laser can be expected in, e.g., the
precise matter-wave interferometry technique [10].

The previous literatures on the coherent PA process are all
focused on the stable and higher atom-molecule conversions by
treating the external optical fields as the classical [4-6, 11].
In this paper, by considering a quantized light as the weak PA
field, we study the quantum dynamics and statistics of coherent
molecular output via two-color PA in an atomic condensate . The
basic physical assumption is that one photon is encoded onto two
atoms to form a diatomic molecule, and then the quantum state of
the photons are converted onto the created stable molecules. As we
described above, the essential features of the current PA
experiments can be well described by the single-mode or even
mean-field approaches, we use the simple single-mode model to
treat this photon-atom-molecule system instead of a fully
analysis. We find two dynamical regimes, the optical transmission
and the molecular formation, and for the later, the quantum state
transfer from light to molecules can be observed, which indicates
a possible way for the quantum control of the molecule-light
system purely by an optical method.

Turning to the situation of Fig. 1, we assume for simplicity that
large number of Bose-condensed atoms are initially prepared in a
magnetic trap when a strong control laser field with slowly
varying Rabi frequency $\Omega(t)$ is applied. The PA is described
by a slowly varying single-mode operator $\hat{a}(t)$, which
induces two atoms to form one quasi-bound molecule in the exited
state $e$. This state mediates the out-coupling Raman transitions
and then the stable molecules are created in the untrapped state
$g$ due to the optical momentum kick [8]. Since the particles
interactions can be tuned by using a Feshbach resonance [12-13]
and goes to zero for a sufficiently $dilute$ condensate, we ignore
it here for the present purpose. In fact, the impacts of nonlinear
particles interactions on the dynamics of the output coherent
matter waves have been well studied in the literatures [14-15]. In
the second quantized notation, boson annihilation operators for
the trapped atoms and the molecules in two states are denoted by
$\hat{b}$, $\hat{e}$ and $\hat{g}$, respectively. By focusing on
the different modes couplings, the effective Hamiltonian of this
photon-atom-molecule system is then ($\hbar=1$)
\begin{equation}
\hat{H}_4=\delta \hat{e}^\dagger \hat{e}+\frac{1}{2}(\epsilon
\hat{e}^\dagger\hat{b}\hat{b}\hat{a}+\epsilon^* \hat{a}^\dagger
\hat{b}^\dagger
\hat{b}^\dagger\hat{e})+[\Omega(t)\hat{g}^\dagger\hat{e}+\Omega^*(t)\hat{e}^\dagger\hat{g}
],
\end{equation}
where $\delta$ is the intermediate detuning and $\hat{a}$ denotes
the quantized input light. Note that we assume here the two-photon
resonance (hence without the $\hat{g}^\dagger \hat{g}$ term) and
also neglect another free-energy term $\hat{a}^\dagger \hat{a}$
since it does not have any essential impact on the main physical
results of our model [9, 24]. This model, comparing with that used
by Winkler $et$ $al.$ to describe their recent PA experiment [4],
only have the new feature of using an input light of quantum
nature, and the incoherent process of the excited-state molecular
damping or the rogue molecular dissociations are also ignored here
[11]. And the weak PA field condition $\epsilon\ll \Omega$ can
safely avoid any heating effects in the PA. Obviously we have:
$\hat{b}^\dagger \hat{b}+2(\hat{e}^\dagger \hat{e}+\hat{g}^\dagger
\hat{g})\equiv N_0$ with $N_0$ being the total number for a
condensate of all atoms or twice the total molecules, indicating
the stationary quantity under the time evolution of the system.

\begin{figure}
\includegraphics[width=8cm]{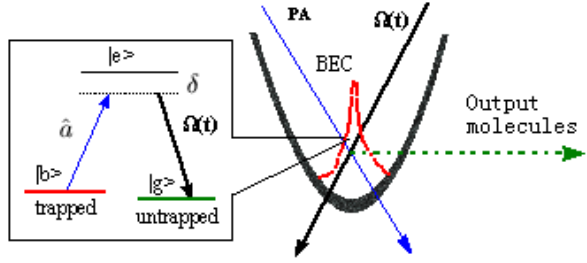}
\caption{(Color online) Schematic figure of the coherent two-color
photo-association by using a quantized input light. The atoms are
initially condensed in the trapped state $|b\rangle$. The
free-quasi-bound-bound transitions and the momentum kick [8] are
induced by two lasers: an input light $\hat{a}(t)$ with Rabi
frequency $\epsilon$ and a classical control light with Rabi
frequency $\Omega(t)$. $\delta$ is the intermediate detuning from
the excited molecular state.} \label{1}
\end{figure}

The roles of atomic association in the two-color PA has been
studied extensively [4-6, 11], even in an optical cavity [16].
Here, by using a quantized input light, we firstly consider the
quantum dynamics of the system under the Bogoliubov approximation,
i.e., by neglecting the atomic depletions and treat $\hat{b}$ as a
$c$-number $b_0$ with $|b_0|^2=N_0$. The general depleted case
will be treated later. It should be mentioned that this atomic
state is assumed to be in a Glauber coherent state instead of a
Fock state. Hence the reduced Hamiltonian describes nothing but a
simple linear three-state system for which the perfect quantum
state transfer is easily expected. In fact, by using the unitary
transformation
\begin{equation}\label{3}
U=\left (
  \begin{array}{ccc}
  \sin\theta_1\sin\theta_2  & \cos\theta_1 & \sin\theta_1\cos\theta_2 \\
  \cos\theta_2 & 0  & -\sin\theta_2 \\
  \cos\theta_1\sin\theta_2  & -\sin\theta_1 & \cos\theta_1\cos\theta_2 \\
  \end{array}
\right ),~~(|U|=-1)
\end{equation}
with the mixing angles defined by
$\tan{\theta_1(t)}=\omega/\Omega(t)$,
$\tan{2\theta_2(t)}=2\lambda(t)/\delta$ and
$\lambda(t)=\sqrt{\omega^2+\Omega^2(t)}$ with $\omega=\epsilon
N_0/2$, the reduced Hamiltonian can be diagonalized as
\begin{eqnarray}\label{4}
\hat{{\mathcal H}_B}&\equiv&U\hat{H}_B
U^\dagger=\omega_+\hat{A}^\dagger\hat{A}+\omega_0\hat{E}^\dagger\hat{E}+
\omega_-\hat{G}^\dagger\hat{G},\nonumber\\
&&\omega_{\pm}=\frac{1}{2}(\delta\pm\sqrt{4\lambda^2(t)+\delta^2});~~~\omega_0=0,
\end{eqnarray}
with $(\hat{a},\hat{e},\hat{g})^T=U(\hat{A},\hat{E},\hat{G})^T$.
Thereby, based on an extending stimulated Raman adiabatic passage
(STIRAP) method [17-18], the quantum transfer process can be
clearly exhibited from the optical state to the molecular state.
And this photon-molecule system then can serve as an another
scenario of the well-known three-state optics (TSO) techniques
which have been intensively studied in recent years for many
systems, e.g., between different internal atomic quantum states
[18], from the light to macroscopic atomic ensembles or
propagating atomic beams [19-20, 13, 21] and vice versa. The new
feature introduced here is that, in some sense, the photons are
stored in the stable diatomic molecules, which may indicate an
optical control of the molecule-light system or even their
entanglement [8].

The essential mechanism of the TSO is the two-channel quantum
interference effect which leads to an effectively unpopulated
intermediate state for certain conditions [18]. This can be
explicitly shown by calculating the three states populations.
Here, by specifying the value of $\Omega$ to make two mixing
angles as time-independent, we find for the intermediate state
\begin{equation}\label{5}
N_e(t)\equiv\langle\hat{e}^\dagger
(t)\hat{e}(t)\rangle=\sin^2\theta_1\sin^2\theta_2\cos^2\theta_2[1-\cos(\omega_+-\omega_-)t]N_0,
\end{equation}
which leads to $N_e(t)=0$ for the conditions of
$(\omega_+-\omega_-)t=2n\pi$ ($n=0,1,2,...$). The detuning
$\delta$ should be large generally in order to avoid the
incoherent processes. Similarly we also can obtain
\begin{equation}\label{6}
N_g(t)=\sin^2\theta_1\cos^2\theta_1[1-\cos(\omega_+t)]N_0.
\end{equation}
Interestingly, we see that under the adiabatic approximations
there can exist two dynamical regimes: (i) for $\omega_+
\tau=2n\pi$, we have an optical transmission regime in which the
coherent PA in fact does $not$ happen ($N_g(\tau)=0$); (ii) only
for $\omega_+ \tau=(2n+1)\pi$, we can observe a coherent molecular
output, with an average flux of particles
\begin{equation}\label{7}
N_g(\tau)=\left[\frac{2\omega\Omega}{\omega^2+\Omega^2}\right]^2N_a(0)\leq
N_a(0).
\end{equation}
And the complete quantum conversion from light to molecules
happens for the conditions of $\omega\approx\Omega$ (or for the
mixing angle, $\theta_1=\pi/4$). This indicates the possibility to
realize a nonclassical molecule laser. Although for a sample of
trapped molecules the photons even can be coherently released
again for the conditions of $\lambda\tau:~(2n+1)\pi\rightarrow
2m\pi$, it is very difficult due to the realistic configurations
in any actual experiment, such as the phase diffusion of trapped
condensate, large inelastic atom-molecule scattering and other
nonideal or dissipated factors especially in a dense condensate.
This is the reason we consider a coherent molecular beam here
instead of memory process in a trapped sample.

Now we proceed to study the quantum statistic of the output
molecular field for an input PA light in a squeezed state [22]:
$|\alpha \rangle_s=\hat{S}(\xi_1)|\alpha\rangle$, where the
squeezed operator $\hat{S}(\xi_1)=\exp[\xi_1
(\hat{a}^{\dag})^2-{\xi}^*_1\hat{a}^2]$ with
$\xi_1=\frac{r_1}{2}e^{-i\phi_1}$, as a unitary transformation on
the Glauber coherent state $|\alpha\rangle
~(\alpha\equiv|\alpha|e^{i\varphi_1})$. The simplest case happens
for the coherent molecular output regime, i.e., $\omega_+
\tau=(2n+1)\pi$, and the Mandel's $Q$ parameter for the output
molecules is obtained as
\begin{equation}\label{8}
Q_g^s(\tau)\equiv\frac{\langle \Delta
\hat{N}_g^2(\tau)\rangle_s}{\langle\hat{N}_g(\tau)\rangle_s}-1
\nonumber\\
=\langle \hat{N}_g(\tau)\rangle_s=4\cos^2\theta_1\sin^2\theta_2
Q_a(0)>0,
\end{equation}
with $Q_a(0)=\sinh^2r_1$ and
$\hat{g}(\tau)=-2\cos\theta_1\sin\theta_1\hat{a}(0)-\cos2\theta_1\hat{g}(0)$.
This indicates a complete quantum conversion (the super-Poisson
distribution) from the input light to the output molecular fields.
In addition, the squeezed angle of the input light also can affect
the quantum statistics of the output molecular field. To see this
clearly, we consider the quadrature squeezed coefficients [23]
\begin{equation}\label{9}
{\cal S}_{ig}=\frac{<(\Delta \hat{{\cal
G}}_i)^2>-\frac{1}{2}|<[\hat{{\cal G}}_1,\hat{{\cal G}}_2]>}
         {\frac{1}{2}|<[\hat{{\cal
G}}_1,\hat{{\cal G}}_2]>|}, \;\; i=1,2
\end{equation}
with $\hat{{\cal G}}_{1}=\frac{1}{2}(\hat{g}+\hat{g}^{\dag})$,
$\hat{{\cal G}}_{2}=\frac{1}{2i}(\hat{g}-\hat{g}^{\dag})$. For the
most interested case: $\omega_+ \tau=(2n+1)\pi$, we can obtain
\begin{equation}\label{10}
 {\cal
 S}^{\xi_a}_{1g,2g}(\tau)=2\sin^2({2\theta_1})\sinh r_1(\sinh r_1\mp \cosh
 r_1\cos\phi_1),
\end{equation}
which indicates a squeezed-angle-dependent molecular squeezed
effect ($r_1>0$): (i) for $\phi_1=2n\pi ~(n=0,1,2,...)$, we have $
{\cal
 S}_{1g}=4\sin^2\theta_1\cos^2\theta_1 (e^{-2r_1}-1)<0$ and $ {\cal
 S}_{2g}=4\sin^2\theta_1\cos^2\theta_1 (e^{2r_1}-1)>0$, which means that
 the component ${\cal
 S}_{1g}$ is squeezed; but for $\phi_1=(2n+1)\pi$, we have $ {\cal
 S}_{1g}={\cal
 S}_{2g}(\phi_1=2n\pi)>0$ and $ {\cal
 S}_{2g}={\cal
 S}_{1g}(\phi_1=2n\pi)<0$, which means that the quantum squeezing
 transfers to ${\cal S}_{1g}$ component; (ii) for $\phi_1=(n+1/2)\pi$,
 we have ${\cal
 S}_{1g}={\cal
 S}_{2g}>0$, a squeezing-free effect for the output molecules [24].

In order to analytically study the impacts of the depletions of
trapped condensate on the coherence of the output molecular field,
we firstly write the Heisenberg equations of motion of the excited
molecules from the Eq.(1) and, by assuming $|\delta|$ as the
largest evolution parameter [11] in the system, we have:
$\hat{e}/\delta=0$, or $
 \hat{e} \simeq
-(\epsilon\hat{b}\hat{b}\hat{a}+\Omega\hat{g})/\delta $.
Substituting this into Eq. (1) to adiabatically eliminate the
quasi-bound molecular state, yields the effective Hamiltonian as
the following
\begin{equation}\label{12}
\hat{H}_3=-\Gamma\hat{g}^\dag\hat{g}-\mu(\hat{a}\hat{a}^\dag\hat{b}\hat{b}\hat{b}^\dag\hat{b}^\dag
+H.c.)-\chi (\hat{g}^\dag\hat{b}\hat{b}\hat{a}+H.c.),
\end{equation}
where $\Gamma=2\Omega^2/\delta$, $\mu=\epsilon^2/\delta$ and
$\chi=2\Omega\epsilon/\delta$. Now we choose $\Omega\gg\epsilon$,
or $\chi\gg\mu$, then it would be good enough to consider the
third term in Eq.(9) as the main couplings of the system. This
observation greatly simplifies our analysis and the Heisenberg
equations of motion for the three different states can be written
as
\begin{equation}\label{13} i \frac{\partial}{\partial
t}\hat{a}=-\chi\hat{b}^\dag\hat{b}^\dag\hat{g}, ~~ i
\frac{\partial}{\partial
t}\hat{b}=-2\chi\hat{a}^\dag\hat{b}^\dag\hat{g}, ~~i
\frac{\partial}{\partial
t}\hat{g}=-\Gamma\hat{g}-\chi\hat{b}\hat{b}\hat{a}.
\end{equation}
The analytical solutions of these equations can be obtained in the
second-order approximation of the evolution time as the following
simple forms:
\begin{eqnarray}\label{14}
\delta\hat{a}(t)&=&it\chi\hat{b}_0^\dag\hat{b}_0^\dag\hat{g}_0+t^2\chi^2\hat{a}_0
(2\hat{N}_{b0}\hat{N}_{g0}-\hat{M}_{b0}), \nonumber\\
\delta\hat{b}(t)&=&2it\chi\hat{a}_0^\dag\hat{b}_0^\dag\hat{g}_0+t^2\chi^2
[\hat{N}_{b0}\hat{b}_0(\hat{N}_{g0}-\hat{N}_{a0})+2\hat{b}_0\hat{N}_{g0}(\hat{N}_{a0}+1)],
\nonumber\\
\delta\hat{g}(t)&=&it\chi\hat{b}_0\hat{b}_0\hat{a}_0-t^2\chi^2\hat{g}_0
[2\hat{N}_{a0}(\hat{N}_{b0}+1)+\hat{M}_{b0}+2\hat{N}_{b0}+1],
\end{eqnarray}
where $\delta\hat{K}(t)\equiv\hat{K}(t)-\hat{K}_0$,
$\hat{N}_{K0}=\hat{K}^\dag_0\hat{K}_0$,
$\hat{M}_{b0}=\hat{b}^\dag_0 \hat{b}^\dag_0 \hat{b}_0\hat{b}_0/2$
and $\hat{K}_0=\hat{a}_0,\hat{b}_0,\hat{g}_0$. Therefore, firstly
by using a coherent factorized structure of the initial state of
the system, i.e., $|in\rangle
=|\alpha\rangle_a|\mu\rangle_b|0\rangle_g$, with
$\hat{b}|\mu\rangle =|\mu|e^{i\varphi_2}|\mu\rangle$, the squeezed
coefficients for the output optical and molecular states are
obtained as
\begin{equation}\label{15}
\left (
  \begin{array}{c}
    S_{1a}(t)\\
    S_{2a}(t)
   \end{array}
\right ) = 3|\alpha|^2(1-|\mu|^2\chi^2t^2)\left (
  \begin{array}{c}
    \cos^2\varphi_1\\
    \sin^2\varphi_1
   \end{array}
\right ),
\end{equation}
and
\begin{equation}\label{16}
\left (
  \begin{array}{c}
    S_{1g}(t)\\
    S_{2g}(t)
   \end{array}
\right ) = 3|\alpha|^2(1-|\alpha|^2|\mu|^2\chi^2t^2)\left [
  \begin{array}{c}
    \sin^2(\varphi_1+2\varphi_2)\\
    \cos^2(\varphi_1+2\varphi_2)
   \end{array}
\right ].
\end{equation}
This means that there is $no$ squeezing for these fields even with
the depletions of the trapped atoms. For the trapped atoms,
however, the dynamical quadrature-squeezed effect always happens
except for the specific case of $|\cos\varphi_2|=|\sin\varphi_2|$.

To examine the quantum state transfer process in the depleted
case, we now use an input light in squeezed vacuum (with nonzero
mean number of photons) [25] for a coherent atomic condensate in
the magnetic trap, and the molecular squeezed coefficients can be
obtained as
\begin{equation}\label{17}
{\cal S}^{\xi_a}_{1g,2g}(t)=2|\mu|^4\chi^2t^2\sinh r_1(\sinh
r_1\mp\cosh r_1\cos\phi_1),~~(\varphi_{1,2}=0)
\end{equation}
from which, comparing with the simplest linear case, we can see
that the squeezed behaviors for the output molecules are also
dependent on the squeezed angle of the input light. However, if
the initial atomic condensate is already prepared in a squeezed
state by, for example, the nonlinear Kerr-type atomic collisions
(see, e.g., Ref.[24]), but the input light is now in an ordinary
Glauber coherent state, i.e., $|in\rangle
=|\alpha\rangle_a|\mu_s\rangle_{b}|0\rangle_g$, $|\mu_s
\rangle_b=\hat{S}(\xi_2)|\mu\rangle_b$, with
$\hat{S}(\xi_2)=\exp[\xi_2
(\hat{b}^{\dag})^2-{\xi}^*_2\hat{b}^2]$, and
$\xi_2=\frac{r_2}{2}e^{-i\phi_2}$, then we get the different
squeezed behaviors for the output molecular field as
\begin{equation}\label{18}
\left (
  \begin{array}{c}
    {\cal S}^{\xi_b}_{1g}(t)\\
   {\cal S}^{\xi_b}_{2g}(t)
   \end{array}
   \right )
= |\alpha|^2\chi^2t^2\sinh^2 r_2\left [ 11\cosh^2 r_2\left (
  \begin{array}{c}
    \sin^2\phi_2\\
    \cos^2\phi_2
   \end{array}
\right)-4 \right ].
\end{equation}
Obviously, the squeezed effects really can appear except for the
case of $|\cos\phi_2|=|\sin\phi_2|$. This may indicate a quantum
transfer from the trapped atoms to the output molecules. However,
it is more difficult for it to be observed in practice since the
dynamical evolutions generally can affect the squeezed parameters
of the atomic field itself (except for some special cases, e.g.,
$|\cos\varphi_2|=|\sin\varphi_2|$), and the initial detection of
the atomic squeezed parameters also can be very challenging for
current experiments of the ultracold atoms . Thereby, it is only
accessible at present to consider the realization of the optical
control of the output coherent molecular waves by using a dilute
trapped condensate.

Finally we can study the mutual coherence of the output optical
and molecular fields by calculating the second-order
cross-correlation function $g_{ag}^{(2)}(t)$ [22]
\begin{equation}
g_{ag}^{(2)}(t)= \frac{\langle \hat{a}^\dag(t) \hat{a}(t)
\hat{g}^\dag(t) \hat{g}(t)\rangle} {\langle \hat{a}^\dag(t)
\hat{a}(t)\rangle \langle \hat{g}^\dag(t) \hat{g}(t)\rangle}.
\end{equation}
Using $\langle \hat{a}^\dag(t) \hat{a}(t) \hat{g}^\dag(t)
\hat{g}(t)\rangle =2\chi^2t^2\langle
\hat{N}_{a0}(\hat{N}_{a0}-1)\hat{M}_{b0}\rangle$, it is easily
verified that this function is independent of the trapped atomic
state and completely determined only by the input optical field.
For the concrete examples, we can obtain the very simple results
$g_{ag}^{(2)}(t)=1-1/|\alpha|^2<1$ (anti-correlated states) for a
coherent input light, and $g_{ag}^{(2)}(t)=3$ (correlated states)
for an input light in squeezed vacuum (with nonzero mean number of
photons, see, e.g., Ref. [25] for the creation and properties of a
light field in squeezed vacuum). This means that both the optical
and molecular fields can exhibit the correlated or anti-correlated
properties with or without the initial optical squeezing. Note
that, these results hold only for the short-time limits and in
this limits, there exists no violation of the Cauchy-Schwarz
inequality (CSI) or nonlocal entanglement of the two output fields
[22]. And it is interesting to further consider the dynamics of
the system in a long time scale and the possibility of creating
entangled molecule-light system via the quantum description of the
two-color PA process.

In conclusion, we theoretically revisit the process of coherent
two-color PA by using a quantized input light. Under the
single-mode approximations, we show that the quantum state
transfer from light to molecules is possible without or with the
atomic depletions. This indicates a possible optical control of
the molecular fields. Of course, the intrinsic molecular collision
in the output beam should be considered, which was shown for an
atom laser also relevant to the squeezing effect [13]. The most
serious simplification here, however, is a single-mode treatment
of the light since our main purpose is about the possible
mechanism of quantum conversion from light to molecules in this
nonlinear system. Although this is suitable for current PA
experiment conditions [4], a multi-mode or spatial structure
analysis can be very significant for any quantum conversion
process, by considering many nonideal factors like the phase
diffusion of the condensate, optical damping and inelastic
two-body scattering. Also, it is difficult to treat the problem of
quantizing a PA light due to the complex microscopic many-body
dynamics [26], and our simple model here should be taken only as
the first step or qualitative probe of the possible features and
applications of this novel nonlinear molecule-light system.

\bigskip
\noindent We are very grateful to the referees for their kind
suggestions which lead to considerable improvement of this paper.
We also thank P. Meystre for his helpful discussions. This work
was supported in part by the National Science Foundation of China
(10304020) and Wuhan Youth Chen-Guang Program.

\end{document}